 \documentclass[aps,prb,twocolumn,showpacs]{revtex4}

\usepackage{graphicx}
\usepackage{color}
\usepackage{units}

\begin{document}

\title{Strongly Anisotropic Magnetic Phase Diagram of CeAu$_{\mathbf{2}}$Ge$_{\mathbf{2}}$}

\author{V.~Fritsch$^1$}
\email[]{veronika.fritsch@kit.edu}
\author{P.~Pfundstein$^2$}
\author{P.~Schweiss$^3$}
\author{E.~Kampert$^4$}
\author{B.~Pilawa$^{1,3}$}
\author{H.~v.~L\"{o}hneysen$^{1,3}$}
\affiliation{$^1$Karlsruher Institut f\"{u}r  Technologie, Physikalisches Institut, 76131 Karlsruhe, Germany \\
             $^2$Karlsruher Institut f\"{u}r  Technologie, Labor f\"{u}r Elektronenmikroskopie, 76131 Karlsruhe, Germany \\
             $^3$Karlsruher Institut f\"{u}r  Technologie, Institut f\"{u}r Festk\"{o}rperphysik, 76131 Karlsruhe, Germany \\
             $^4$Helmholtz-Zentrum Dresden-Rossendorf, Hochfeld-Magnetlabor Dresden, 01314 Dresden, Germany}

\date{\today}

\begin{abstract}
CeAu$_2$Ge$_2$ single crystals (tetragonal ThCr$_2$Si$_2$ structure) have been grown in Au-Ge flux (AGF) as well as in Sn flux (SF). X-ray powder-diffraction and EDX measurements indicate that in the latter case Sn atoms from the flux are incorporated in the samples, leading to a decrease of the lattice constants by  $\approx 0.3 \%$ compared to AGF samples. For both types of samples, a strong anisotropy of the magnetization $M$ for the magnetic field $\mathbf{B}$ parallel and perpendicular  to the $\mathbf{c}$ direction is observed with $M_{||}/M_{\perp} \approx 6 - 7$ in low fields just above $10$~K. This anisotropy is preserved to high fields and temperatures and can be quantitatively explained by crystal electric field effects.
Antiferromagnetic ordering sets in around $10$~K as previously found for polycrystalline samples.  From the magnetization data of our single crystals we obtain the phase diagrams for the AGF and SF samples. The magnetic properties depend strongly on the flux employed. While the AGF samples exhibit a complex behavior indicative of several magnetic transitions, the SF samples adopt a simpler antiferromagnetic structure with a single spin-flop transition. This effect of a more ordered state induced by disorder in form of Sn impurities is qualitatively explained within the ANNNI model, which assumes ferromagnetic and antiferromagnetic interactions in agreement with the magnetic structure previously inferred from neutron-scattering experiments on polycrystalline CeAu$_2$Ge$_2$ by Loidl {\it et al.} [Phys. Rev. B {\bf 46}, 9341, (1992)].
 \end{abstract}


\maketitle

\section{Introduction}
Intermetallic compounds with ThCr$_2$Si$_2$ structure have been studied in depth in the past revealing new phenomena in different areas,
e.g., unusual quantum criticality in YbRh$_2$Si$_2$ (Ref.~\onlinecite{custers2003}) or high-temperature superconductivity in Fe pnictides.\cite{rotter2008} Further, there is the evolution of spin-glass behavior in the series PrAu$_2$(Ge$_{1-x}$Si$_x$)$_2$ (Ref.~\onlinecite{krimmel1999}), whose origin is still under debate.\cite{goremychkin2007,goremychkin2008} In particular the Ce$M_2$Ge$_2$ and Ce$M_2$Si$_2$ compounds, where $M$ is a noble metal, exhibit a broad variety of ground states: The first heavy-fermion superconductor ever discovered was CeCu$_2$Si$_2$ (Ref.~\onlinecite{steglich1979}). As opposed to the paramagnets CePt$_2$Si$_2$ and CeRu$_2$Si$_2$, where Ce is intermediate valent,\cite{ayache1987,gupta1983} in Ce$M_2$Ge$_2$ and Ce$M_2$Si$_2$ with $M =$ Ag or Au, the Ce ions adopt a nearly trivalent state, resulting in long-range antiferromagnetic order.\cite{severing1989,loidl1992} The exact nature of the antiferromagnetic ground state in these systems, if present, e.g., whether it is commensurate or incommensurate, helical, with the spins along the $\mathbf{a}$ or $\mathbf{c}$ axis, is determined by the complex interplay of crystal structure and the electronic structure, RKKY interactions and crystal electric fields, all of which may lead to pronounced anisotropies in the magnetic properties.

In this paper we will focus on CeAu$_2$Ge$_2$. Despite the anisotropy of the magnetic structure most previous investigations on CeAu$_2$Ge$_2$ were performed on polycrystals, where a N\'{e}el temperature of $16$~K was reported.\cite{loidl1992} In order to explore the role of anisotropy, several groups have recently grown single crystals by flux growth methods, e.g., CeAu$_2$Si$_2$ (Ref.~\onlinecite{sefat2008}) and CeAg$_2$Ge$_2$ (Ref.~\onlinecite{thamizhavel2007}), as well as CeAu$_2$Ge$_2$ (Ref.~\onlinecite{joshi2010}). Joshi {\it et al.} prepared single crystals of CeAu$_2$Ge$_2$ from Bi flux and found a strong uniaxial anisotropy.\cite{joshi2010} In this work we used two different flux materials (Sn vs. Au-Ge). Our investigations reveal a strong dependence of the magnetic properties on the flux employed. We discuss the magnetic phase diagram of CeAu$_2$Ge$_2$ determined from magnetization measurements, which crucially depends on magnetic anisotropy and crystal field effects.

\section{Experimental details and sample characterization}
Single crystals were grown from high-purity starting materials\cite{ames} by a flux growth method in Au-Ge flux (AGF), utilizing the eutectic point in the binary Au-Ge phase diagram at $361^{\circ}$C, similar as described elsewhere.\cite{thamizhavel2007,sefat2008} Another set of samples was grown in Sn flux (SF). The samples grow in plate-like shapes of about half a mm thickness and an area of several mm$^2$.

\begin{figure}
\includegraphics[clip,width=0.49\linewidth]{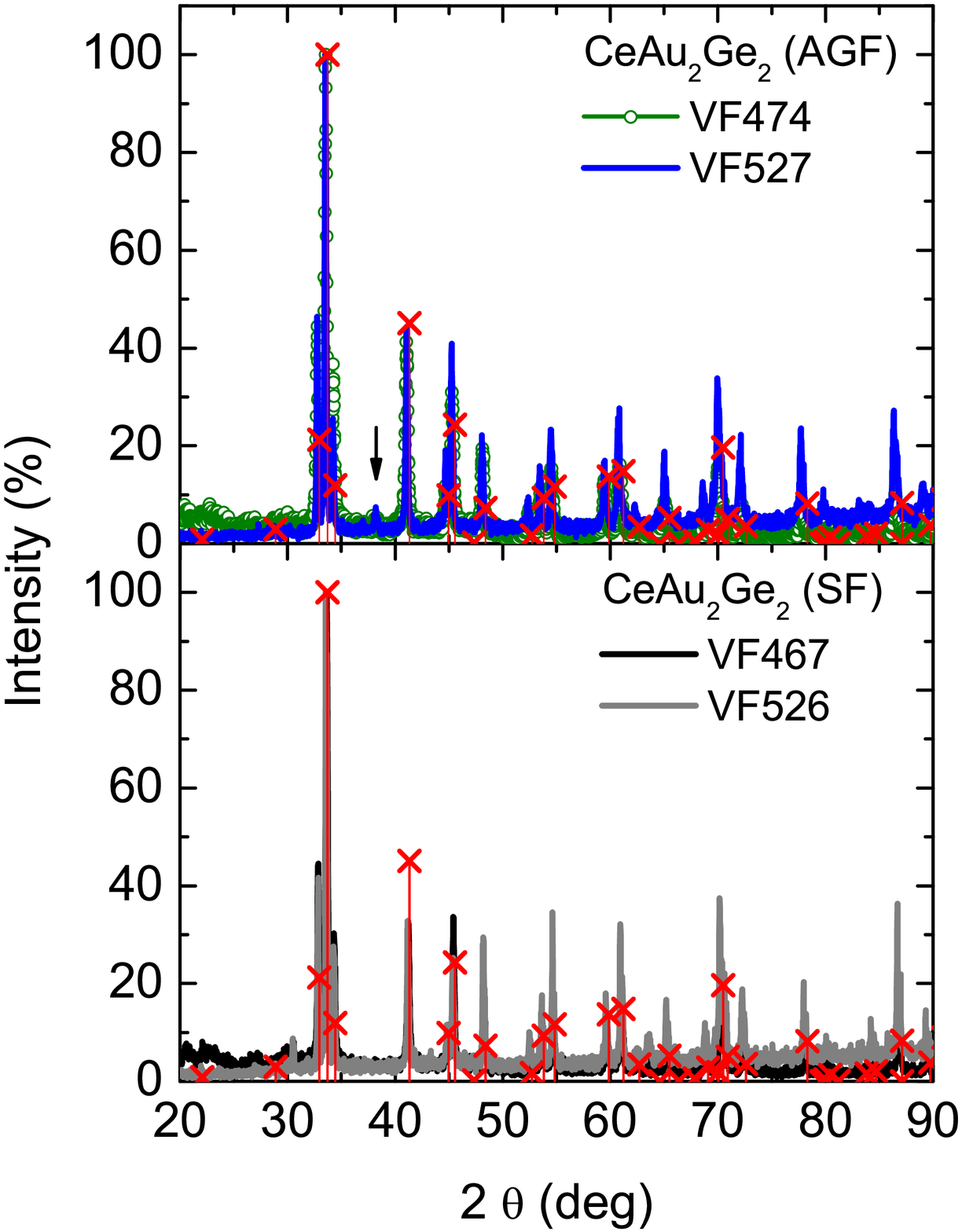}
\includegraphics[clip,width=0.49\linewidth]{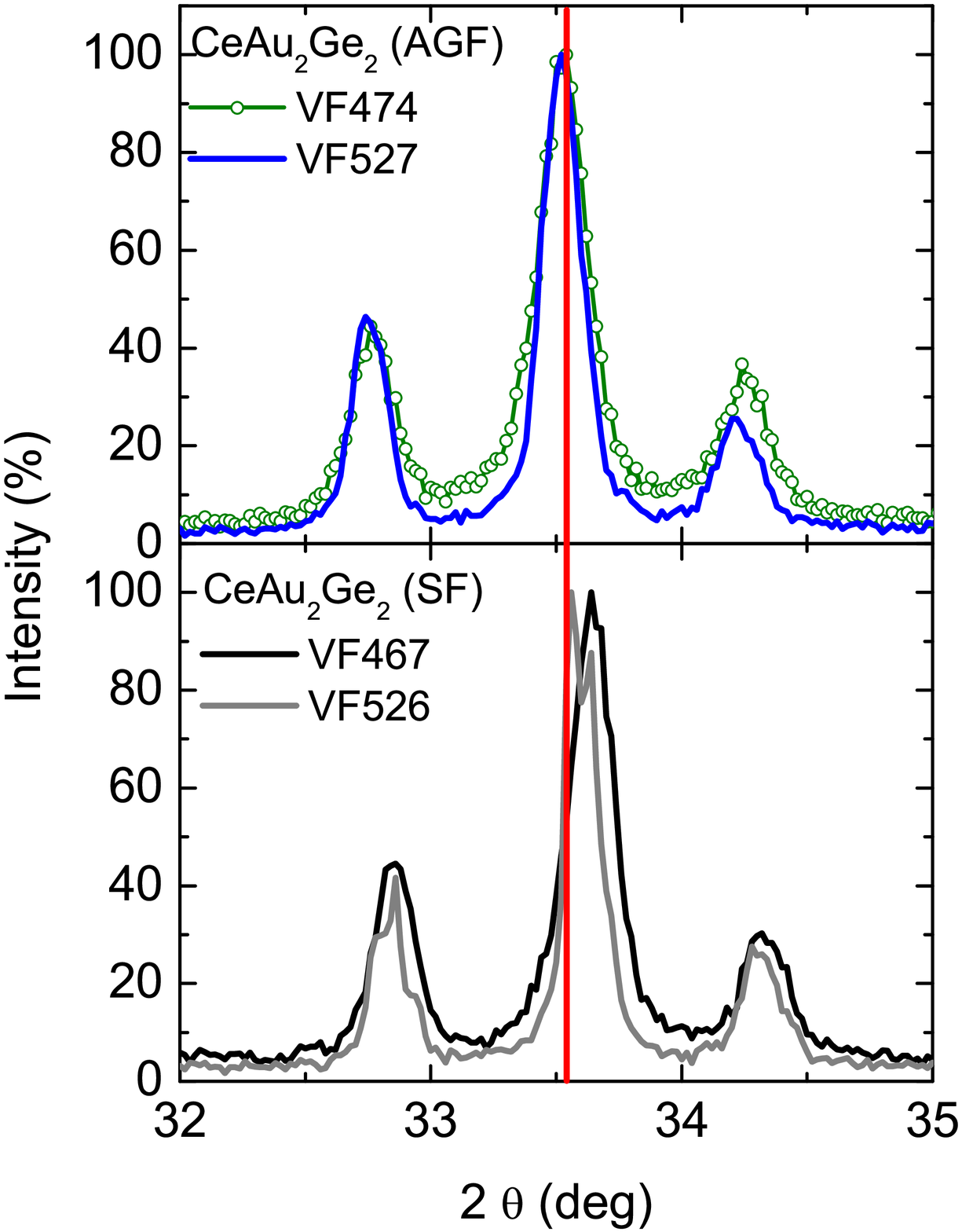}
\caption{(Color online) Left: X-ray powder diffraction patterns of several samples of CeAu$_2$Ge$_2$ grown in Au-Ge or Sn flux (denoted by AGF and SF, respectively). The crosses mark the calculated peak heights and positions.  The arrow in the top panel indicates Au inclusions. Right: ($103$), ($112$), and ($004$) peaks on an expanded horizontal scale. The vertical line illustrates the systematic shift of the center peak between AGF and SF samples.} \label{fritsch_f1}
\end{figure}

\begin{table}
\begin{ruledtabular}
\begin{tabular}{lcc}
& $a$~(\AA) & $c$ (\AA)  \\
\hline
AGF VF527 & 4.39248 & 10.4736  \\
AGF VF474 & 4.39233 &  10.4654 \\
SF VF467 & 4.38119 & 10.4482  \\
SF VF526 & 4.38026  & 10.4446   \\
\end{tabular}
\caption{Lattice parameters of several CeAu$_2$Ge$_2$ samples derived from x-ray powder diffraction patterns.} \label{fritsch_t1}
\end{ruledtabular}
\end{table}

All samples were characterized by x-ray powder diffraction. The diffraction patterns confirmed the samples to be single phase with ThCr$_2$Si$_2$ structure, space group No. 139. The powder diffraction patterns of the AGF and SF samples are shown in Fig.~\ref{fritsch_f1}. The right-hand panel  reveals a clear shift of peak positions between AGF and SF samples, indicating that the lattice parameters of the SF samples are somewhat smaller than those of the AGF samples (see also Tab.~\ref{fritsch_t1}).

\begin{table}
\begin{ruledtabular}
\begin{tabular}{lcccc}
&  at$\%$ Ce & at$\%$ Au & at$\%$ Ge & at$\%$ Sn \\
\hline
AGF VF527   & 20.45  & 39.59 & 39.96  & -\\
AGF VF474   & 20.66 & 39.51 & 39.83 & - \\
SF VF467  & 20.92 & 36.25 & 39.11 & 3.73  \\
SF VF526  & 20.38  & 38.76  & 36.62 &4.25 \\
\end{tabular}
\caption{Comparison of CeAu$_2$Ge$_2$ determined with EDX.} \label{fritsch_t2}
\end{ruledtabular}
\end{table}

\begin{figure}
\includegraphics[clip,angle=-90,width=0.49\linewidth]{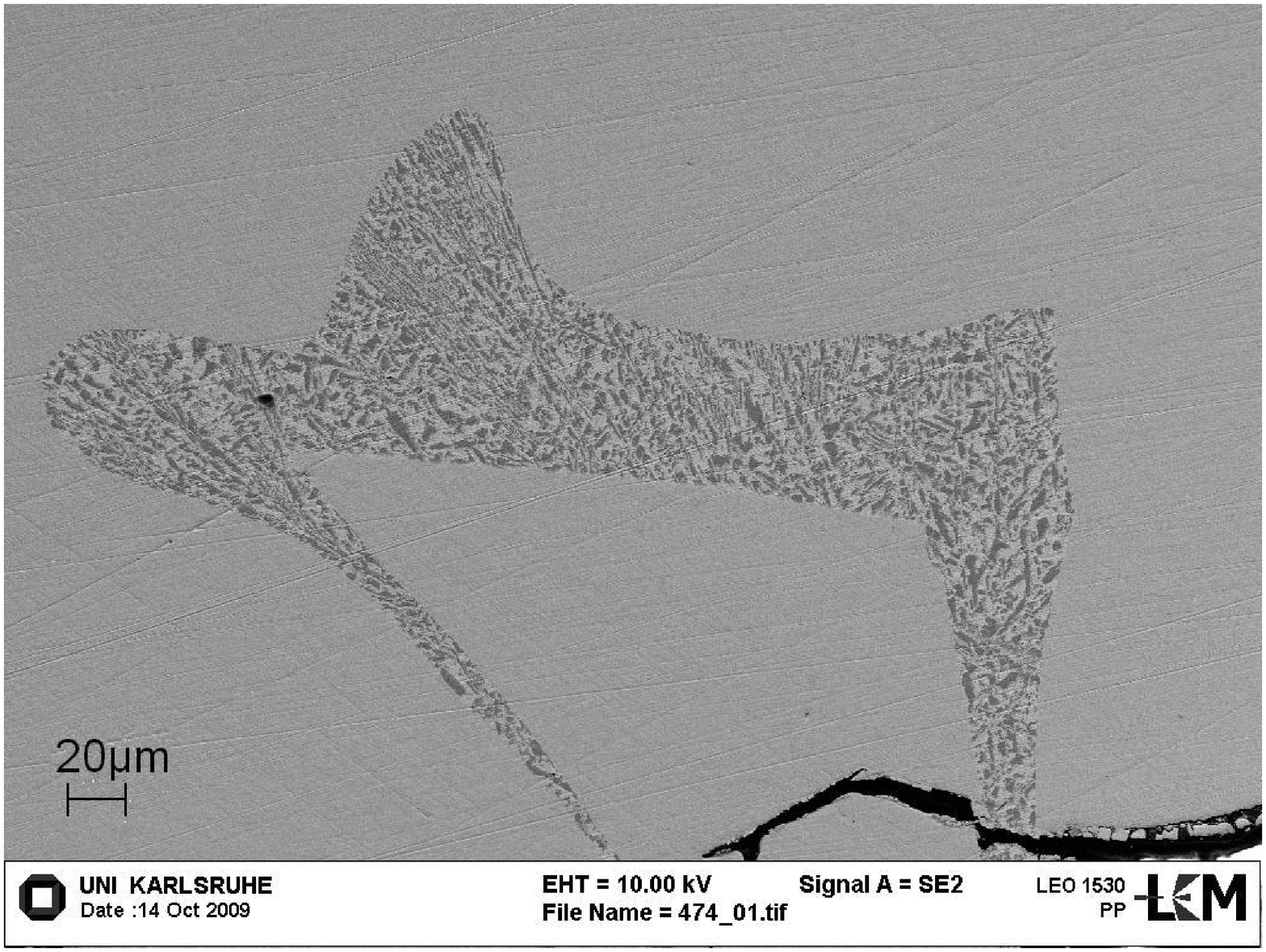}
\includegraphics[clip,angle=-90,width=0.49\linewidth]{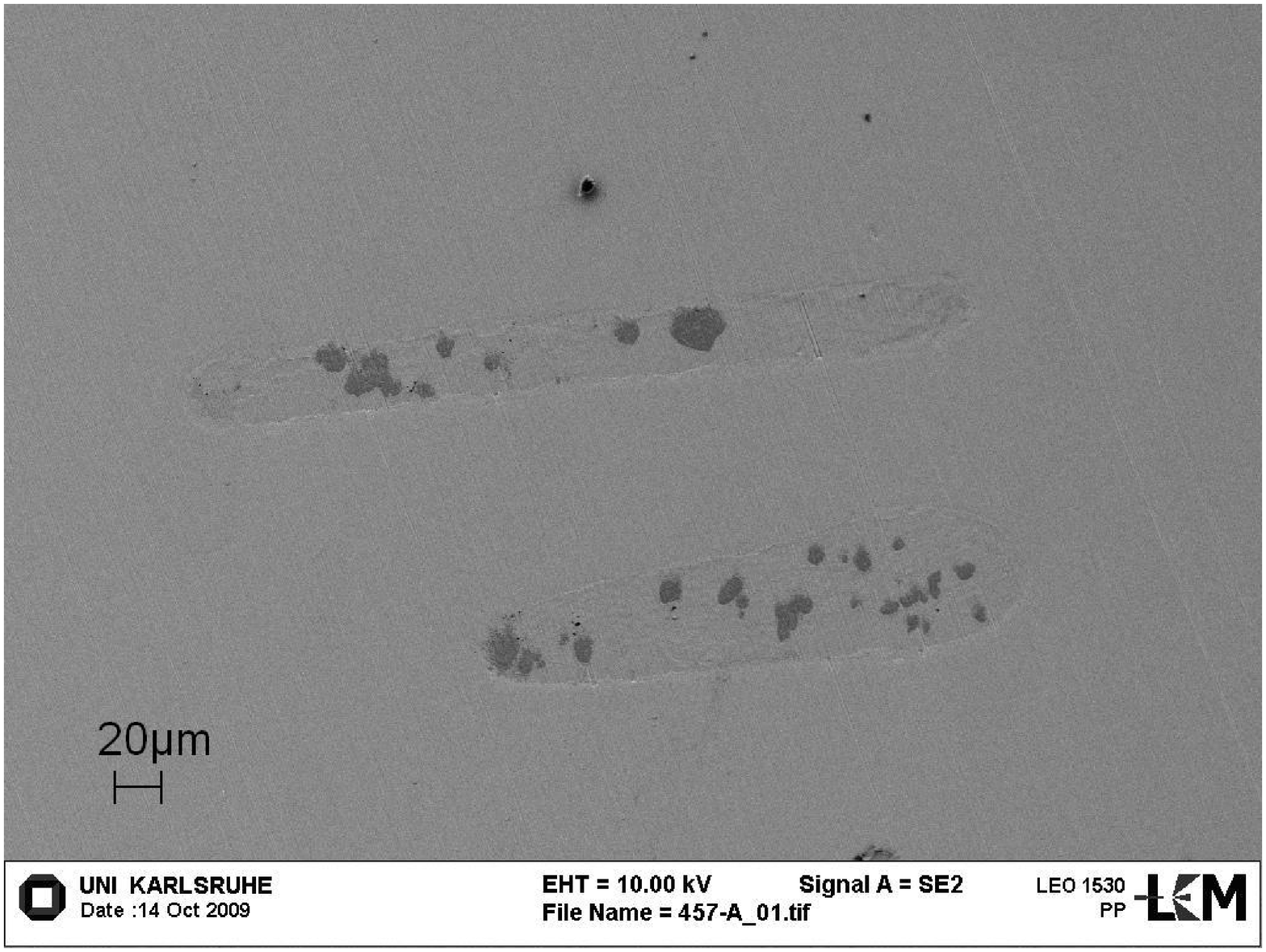}
\caption{Electron microscopy pictures, left: sample VF474 AGF, right: sample VF467 SF (see text for details)} \label{fritsch_f2}
\end{figure}

We inspected the samples with scanning electron microscopy. In addition to the homogeneous majority phase, a few light and dark inclusions were found, as shown in Fig.~\ref{fritsch_f2}. In the AGF samples these inclusions were identified by EDX measurements as pure Au (light) and pure Ge (dark), in the SF samples the inclusions turned out to be Au-Sn and Ge-Sn alloys. Pure Au is also visible as a tiny peak in the x-ray patterns of the AGF samples, as indicated by a small arrow in the upper left-hand panel of Fig.~\ref{fritsch_f1}. We estimate the total volume fractions of foreign inclusions to $\lesssim 3 \%$, as inferred from the nearby complete absence of inclusion-related x-ray peaks.

Apart from these inclusions, the stoichiometry of the AGF samples closely resembles the nominal composition, while the bulk of the SF samples contains about $4$ at$\%$  Sn (see Tab.~\ref{fritsch_t2}). Since no Ce is incorporated in these inclusions, the inclusions are not expected to affect the magnetic properties of the samples.

A single crystal of each batch was inspected by four-circle x-ray diffractometry to determine the structural parameters. The results are summarized in Tab.~\ref{fritsch_t3}. They confirm the lattice parameters of the SF samples to be smaller than those of the AGF samples. The occupancy of the Au site increases with increasing lattice parameter. With higher Au content the  Au-Ge bond length grows, together with the distance of the Au-Ge layers. These findings corroborate the assumption that in the SF samples Sn is incorporated  on the Au sites. However, one should note that the distance of the Ce-Ge layers is slightly reduced in the AGF samples compared to the SF samples.

\begin{table*}
\begin{ruledtabular}
\begin{tabular}{lccccc}
             & bond length &\multicolumn{2}{c}{layer distance} & \multicolumn{2}{c}{occupancy}  \\
          	 &  Au-Ge	&  Au-Ge	&  Ce-Ge	&  Au	&  Ge	\\
\hline
AGF VF527     &	2.6082	& 2.8072	& 2.4403 & 0.968	& 1.04  \\
AGF VF474		 & 2.605	& 2.8016	& 2.4374 & 0.957	& 1.05  \\
SF VF467	     & 2.5916  & 2.7718	& 2.4572 & 0.920	& 1 \\
SF VF526	  & 2.5915	& 2.7661	& 2.4577 & 0.918	& 1.02  \\
\end{tabular}
\caption{Single-crystal x-ray results for several samples of CeAu$_2$Ge$_2$. The occupancies are normalized to $100\,\%$ Ce. } \label{fritsch_t3}
\end{ruledtabular}
\end{table*}

The magnetization $M$ was measured with a commercial vibrating-sample magnetometer (VSM) from Oxford Instruments in the temperature range between $1.6$ and $300$~K in magnetic fields $B$ up to $12$~T. The magnetization measurements around $\approx 2$~K in fields up to $60$~T were carried out at the Hochfeld-Magnetlabor Dresden in a pulsed magnet with $20$~mm bore with an inductive coil system, as described elsewhere.\cite{zherlitsyn2010}  A digitizer recorded the signal of the pick-up coils, which was then integrated numerically. The signal of the field up-sweeps was affected by a spike at the beginning of the pulse and therefore corrected by a constant off-set extracted from the down-sweeps.  The average of multiple sweeps obtained for each sample was calibrated with the VSM data. While this procedure worked well for the SF sample, the AGF sample exhibits a noteworthy difference in curvature between the low-field VSM data and the pulsed-field data, which could not be resolved. The magnetic field was determined by an additional set of pick-up coils calibrated by measuring the magnetization of MnF$_2$, which exhibits a well-known spin-flop transition at $B = 9.27$~T.\cite{felcher1996,skourski2011}

Preliminary resistance measurements have shown a room-temperature resistivity of about $50~\mu\Omega$cm for the AGF samples, for the SF samples the values vary between $65$ and $95~\mu\Omega$cm.  However, the residual resistance ratio is between $2$ and $3$ for all samples, hence they are poor metals. These preliminary measurements of $\rho(T)$ show a kink-like minimum at the N\'{e}el temperature that is used to corroborate the $B$-$T$ phase diagram.

\section{Results}
\begin{figure}
\includegraphics[clip,width=0.49\linewidth]{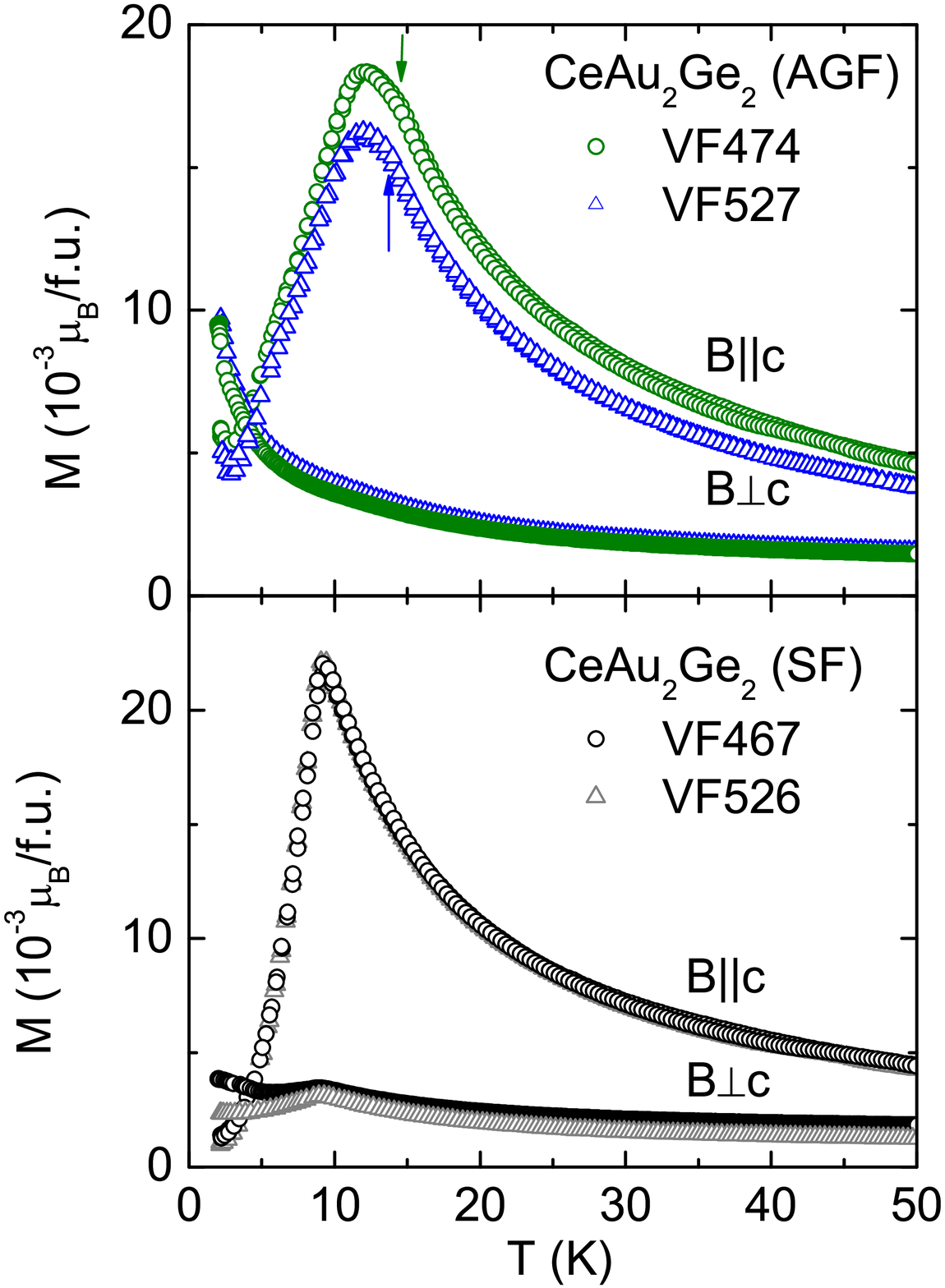}
\includegraphics[clip,width=0.465\linewidth]{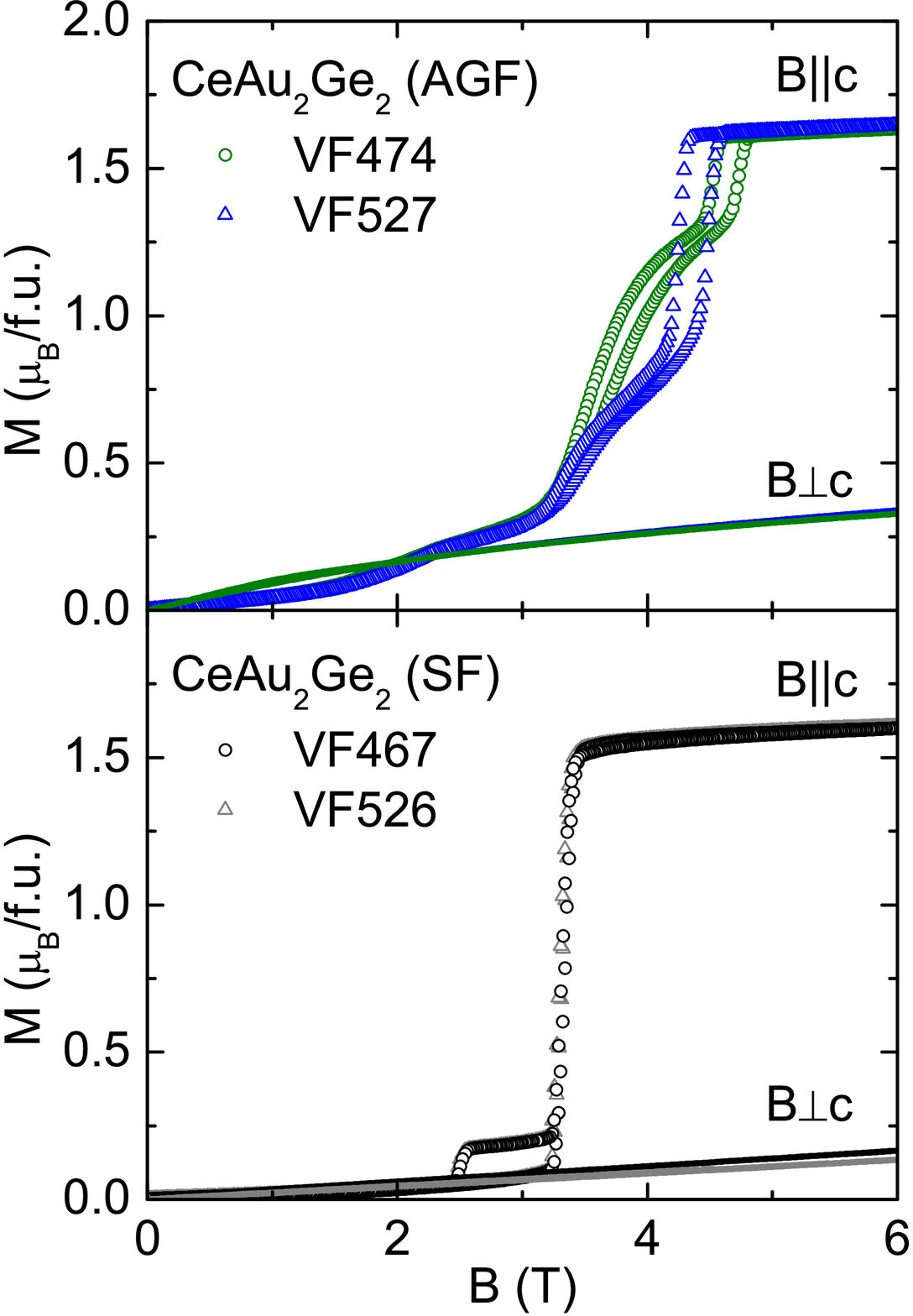}
\caption{(Color online) Left: Magnetization $M$ vs temperature $T$ of AGF and SF CeAu$_2$Ge$_2$  with the magnetic field $B = 0.1$~T aligned either parallel or perpendicular to the $\mathbf{c}$ axis. The arrows in the upper frame mark additional anomalies in the magnetization curves (see main text). Right: Magnetization $M$ vs. external magnetic field $B$ at $T = 2.6$~K of AGF and SF CeAu$_2$Ge$_2$.} \label{fritsch_f3}
\end{figure}

The magnetic properties of CeAu$_2$Ge$_2$  at low temperatures differ substantially, due to the use of different fluxes but are reproducible for different batches of the same flux, as will be discussed in detail below. As a common feature, all samples show a strong magnetic anisotropy with the $\mathbf{c}$ axis being the easy direction. The left-hand panel of Fig.~\ref{fritsch_f3} displays the temperature dependence of the magnetization $M$ in an external magnetic field $B = 0.1$~T. The magnetization $M$ shows a sharp peak at $12$~K (AGF) or $9$~K (SF) for ${\mathbf B}  \| {\mathbf c}$, but for  ${\mathbf B}  \bot {\mathbf c}$ no anomaly is seen in the AGF samples and a very shallow maximum only in the SF samples. For ${\mathbf B}  \| {\mathbf c}$ the AGF samples show an upturn of $M$ below $3$~K  which is absent for SF samples, hinting at a more complex magnetic structure in the former. No differences between zero-field-cooled and field-cooled data are found in any sample. Additionally an anomaly (marked by arrows) in the AGF samples is found at $T \approx 15$~K, close to the value for polycrystalline samples reported by Loidl {\it et al.}\cite{loidl1992}

The right-hand panel of Fig.~\ref{fritsch_f3} shows the field-dependent magnetization $M$ at $T = 2.6$~K. The AGF samples undergo for $\mathbf{B} \| \mathbf{c}$ several broad metamagnetic transitions with a sizable hysteresis, indicating first-order transitions. Upon further increasing the field, the magnetization continues to rise weakly reaching $M(12~\unit{T}) = 1.7 \mu_B/$f.u. in the two samples at $B = 12$~T (see Fig.~\ref{fritsch_f4} below). The SF samples exhibit only one, albeit very sharp, metamagnetic transition, as opposed to the more complex magnetic structure in the AGF samples. Upon decreasing field, a hysteretic tail is observed.

\begin{figure}
\includegraphics[clip,width=0.49\linewidth]{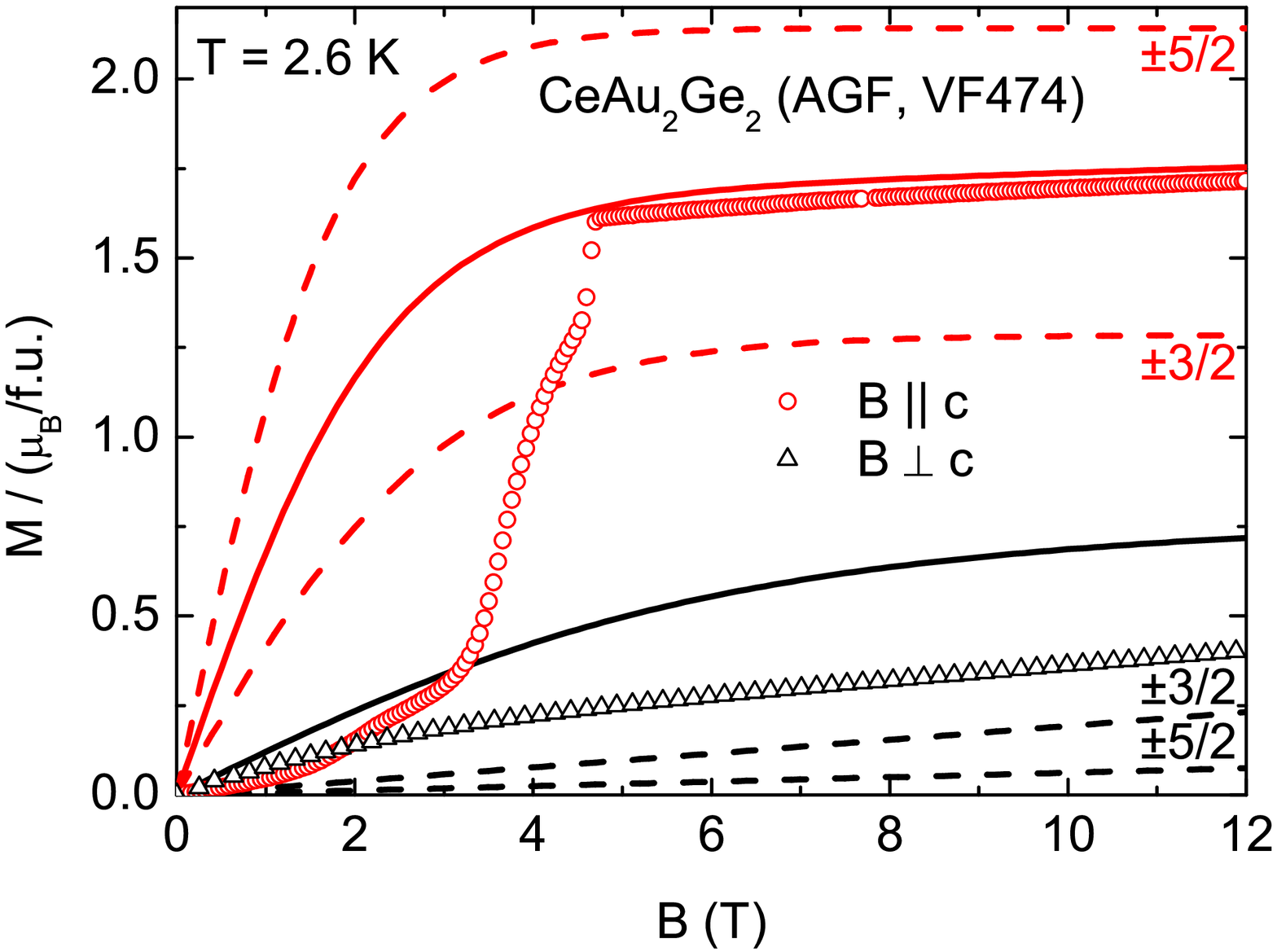}
\includegraphics[clip,width=0.49\linewidth]{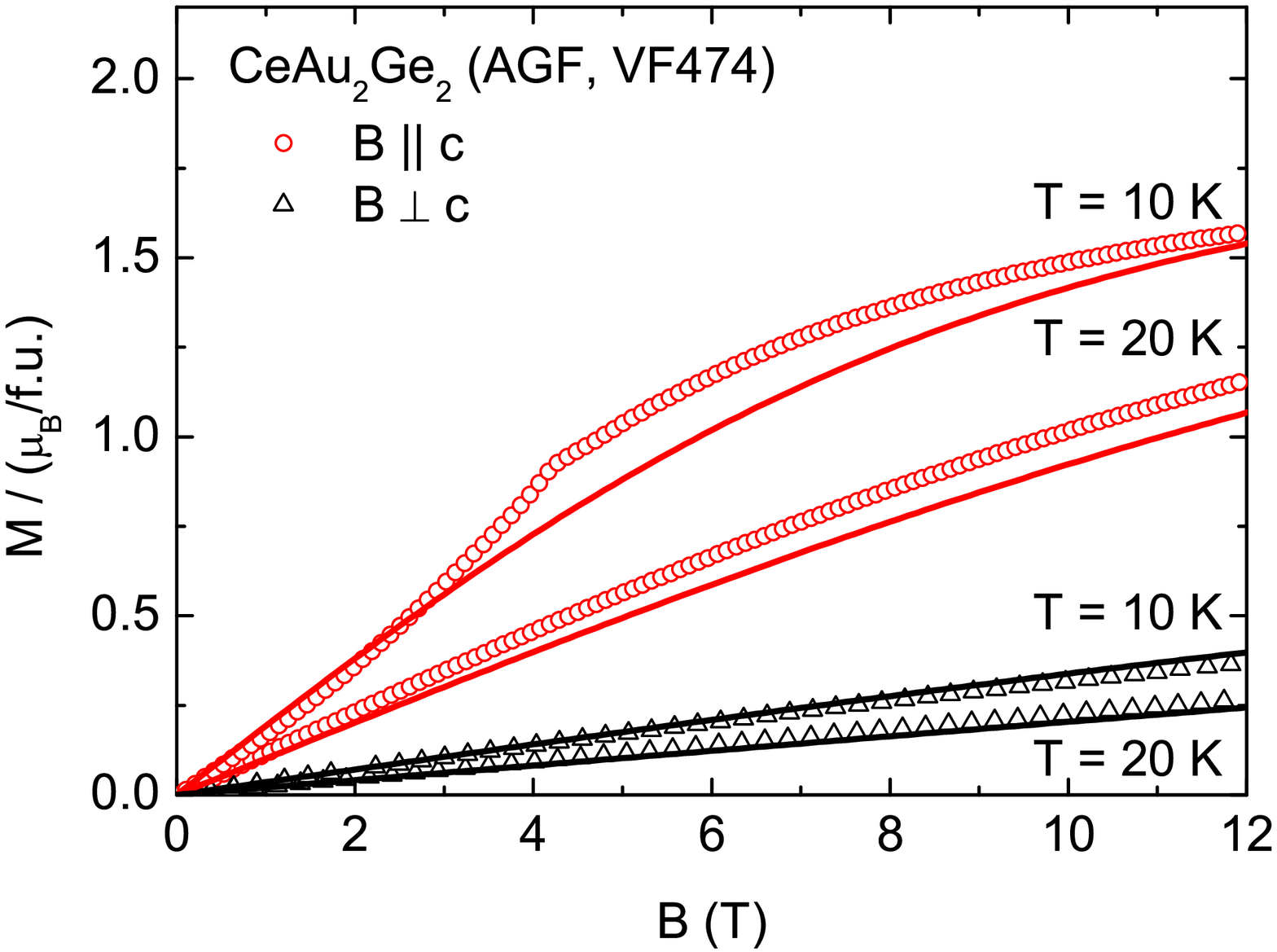}
\caption{(Color online) Comparison of measured and calculated magnetization of AGF CeAu$_2$Ge$_2$. For details see text. Left: Measured and calculated data for $T = 2.6$~K. The solid line is the calculated magnetization with the crystal-field parameters from ref.~\onlinecite{loidl1992}, the dashed lines indicate the magnetization for the pure $\left|\pm\frac{5}{2}\right>$  and $\left|\mp\frac{3}{2}\right>$  states.  Right: Calculated and measured data for $T = 10$ and $20$~K. For clarity, only the data taken with increasing field are shown. }\label{fritsch_f4}
\end{figure}

We calculated the single-ion magnetic properties of  CeAu$_2$Ge$_2$ on the basis of the crystal-field analysis of inelastic neutron-scattering experiments by Loidl {\it et al.}\cite{loidl1992} In CeAu$_2$Ge$_2$ the two Ce sites are equivalent and of tetragonal site symmetry. The crystal-field operator is a linear combination of the operators $O_2^0$, $O_4^0$ and $O_4^4$ with the coefficients  $B_2^0$, $B_4^0$ and $B_4^4$. The $J = \frac{5}{2}$ ground multiplet of the Ce$^{3+}$ ion is split into three doublets. The ground state is a nearly pure $\left|\pm\frac{5}{2}\right>$  state and the first excited level a pure $\left|\pm \frac{1}{2}\right>$ state. The excitation energies of the two excited states $E_1$ and $E_2$, the mixing angle between the $\left|\pm\frac{5}{2}\right>$  and $\left|\mp\frac{3}{2}\right>$  states and the crystal-field coefficients have been given by Loidl et al.\cite{loidl1992} To calculate the field dependence of the magnetization and the temperature dependence of the magnetic susceptibility, we use in the parameters $E_1=11$~meV, $E_2=17.2$~meV and the mixing angle of $22.2^{\rm o}$ (ref.~\onlinecite{loidl1992}). The calculated $M(B)$ curves are shown in Fig.~\ref{fritsch_f4} together with the experimental data of AGF CeAu$_2$Ge$_2$ for $T = 2.6$~K (in the magnetically ordered regime), $T = 10$~K (close to the magnetic phase transition) and $T = 20$~K (in the paramagnetic regime).

At high temperatures $T = 20$~K in the paramagnetic regime, the calculated $M(B)$ curves coincide fairly well with our experimental data for both field directions, as shown in the right-hand panel of Fig.~\ref{fritsch_f4}. As expected, the magnetization curve cannot be described within a single-ion model at lower temperatures where cooperative effects gain importance: at $T = 10$~K in low fields $\mathbf{B} \| \mathbf{c}$ one can see clear deviations from the calculated behavior, in high fields measured and calculated data converge again. At $2.6$~K the calculated single-ion magnetization is reached at high fields after the metamagnetic transitions.

\begin{figure}
\includegraphics[clip,width=0.49\linewidth]{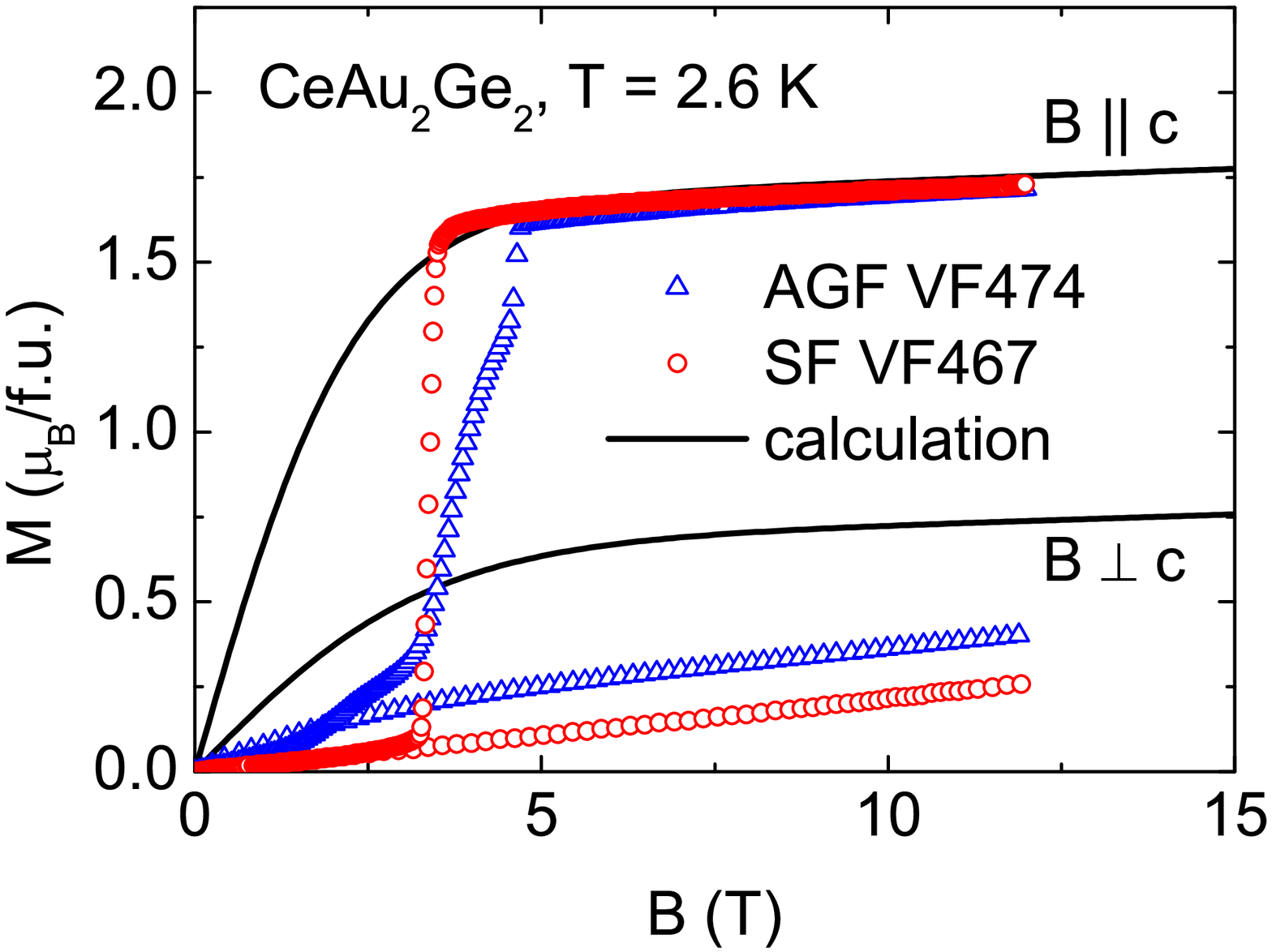}
\includegraphics[clip,width=0.49\linewidth]{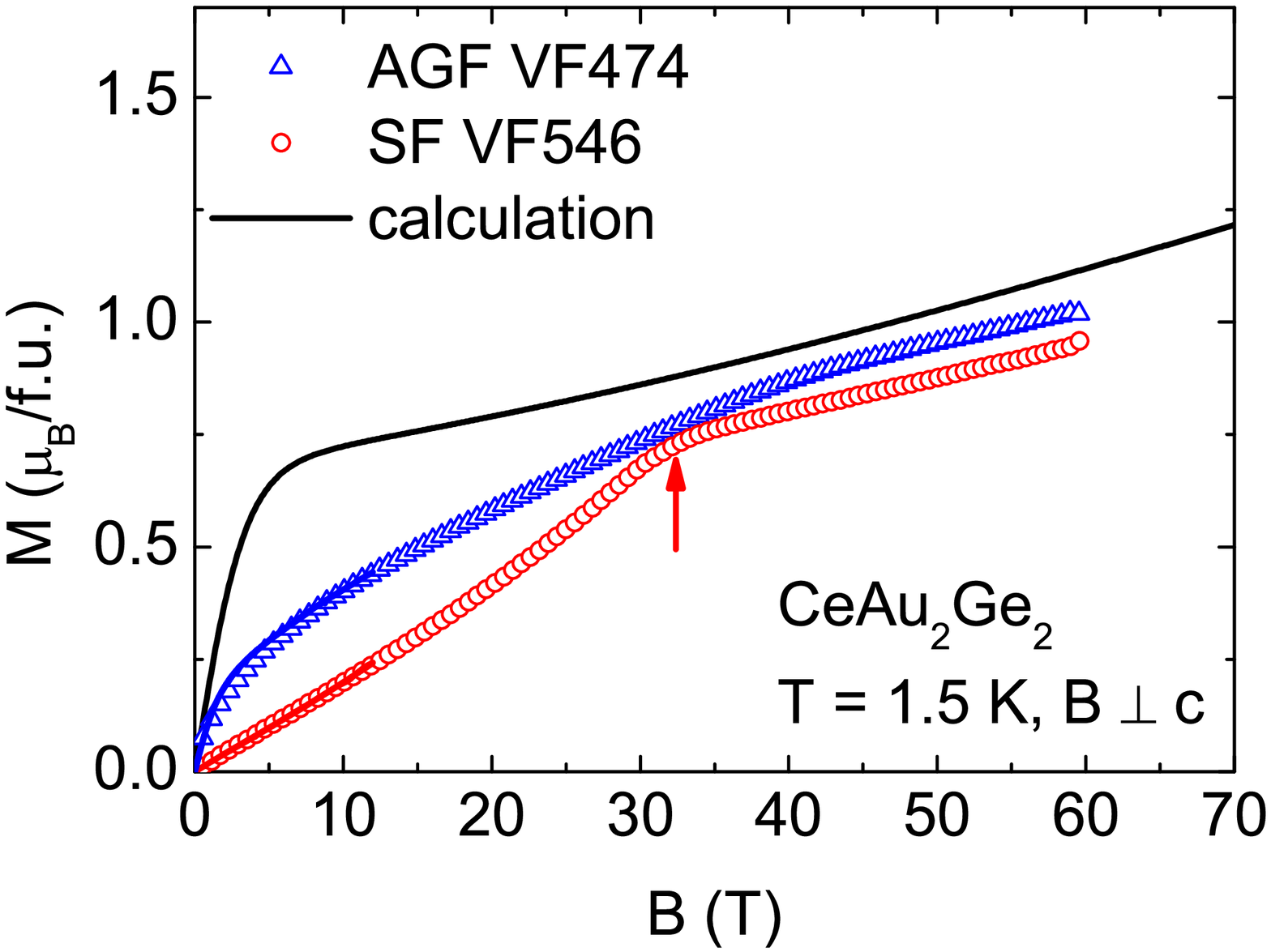}
\caption{(Color online)  Left: Magnetization $M$ of SF (red circles) and AGF (blue triangles) grown CeAu$_2$Ge$_2$ compared to the calculated data in fields up to $15$~T with the magnetic field $B$ aligned parallel and perpendicular to the $c$-axis.  Right:  Magnetization $M$ of SF (red circles) and AGF (blue triangles) grown CeAu$_2$Ge$_2$ compared to the calculated data (solid black line) in fields up to $60$~T with the magnetic field $B$ aligned perpendicular to the $c$-axis. The solid lines below $12$~T indicate the low-field VSM data.}\label{fritsch_f5}
\end{figure}

However, for $\mathbf{B} \perp \mathbf{c}$ the experimental data fall short of the calculated curve by a large margin. Moreover, there is a distinct difference between SF and AGF samples, as shown in the left-hand panel of Fig.~\ref{fritsch_f5}: at $12$~T the magnetization $M$ of the SF sample amounts to only $60\,\%$ of the value of the AGF sample. In order to find the origin of these strong deviations from the calculated behavior, we performed magnetization measurements in fields $\mathbf{B} \perp \mathbf{c}$ up to $60$~T, which are shown in the right-hand panel of Fig.~\ref{fritsch_f5}.

The magnetization of the AGF sample increases smoothly with a downward curvature in the whole field range, approaching the calculated value  at approximately $40$~T and then grows further closely following the calculated $M(B)$ curve. In contrast, the magnetization of the SF sample increases slightly superlinearly with an upward curvature at low fields, and exhibits a distinct kink at $B^{\ast} = 32$~T, marked by the red arrow, and then increases further. At high fields the data agree fairly well with the calculated behavior. The small differences between the measured data of the two samples at the highest fields may be attributed to experimental errors in the exact calibration of the magnetometer. These might also account for the difference from the calculated curve, with the additional uncertainty in the determination of the mixing angle.

\begin{figure}
\includegraphics[clip,width=0.49\linewidth]{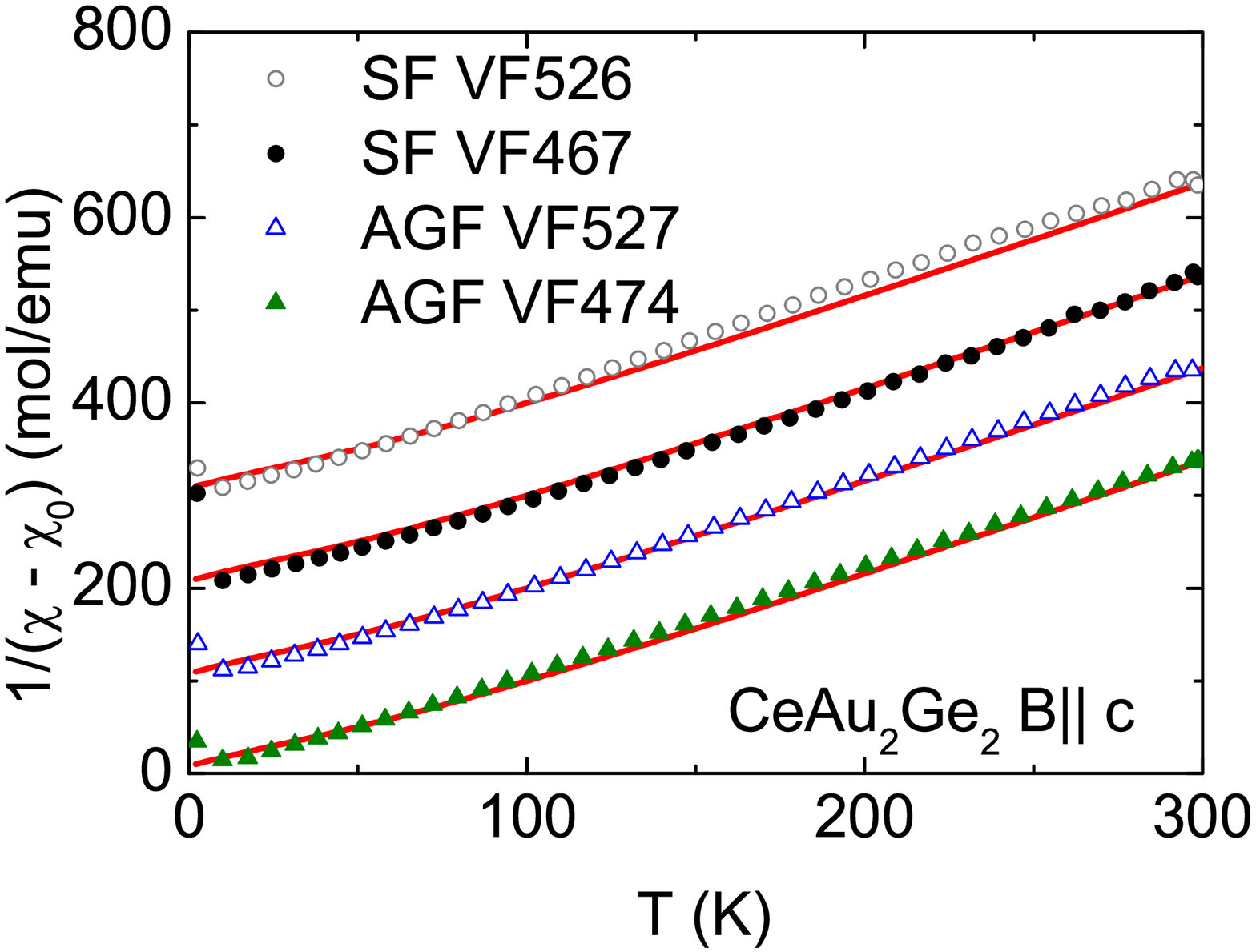}
\includegraphics[clip,width=0.49\linewidth]{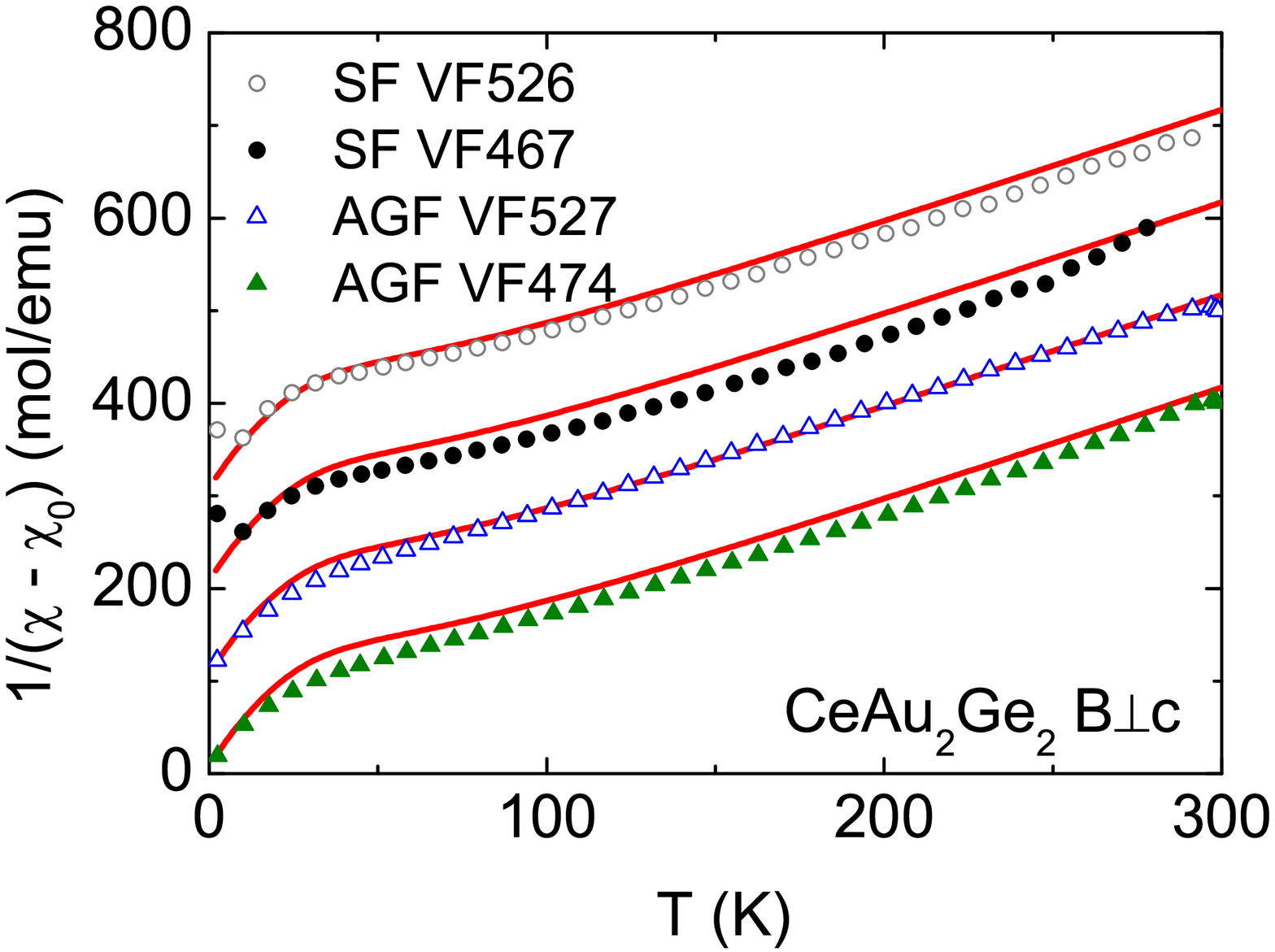}
\caption{(Color online) Inverse dc susceptibility $\chi - \chi_0$ vs. temperature of CeAu$_2$Ge$_2$ in an external field $B = 100$~mT aligned $\mathbf{B} \| \mathbf{c}$ (left) and $\mathbf{B} \bot \mathbf{c}$ (right). The solid red lines are calculated susceptibilities on the basis of the neutron scattering data of Loidl et al.\cite{loidl1992}. For the determination of $\chi_0$ see text. For clarity the data are shifted by $100\,\unit{mol/emu}$ each.}\label{fritsch_f6}
\end{figure}

The magnetic susceptibility $\chi$ measured in low fields $B = 0.1$~T can be described taking into account the free-ion susceptibility of the $^2F_{5/2}$ ground state modified by the crystal-electric-field splitting\cite{loidl1992} with an additional mean-field interaction constant $\lambda$ (ref.~\onlinecite{joshi2010}), after subtracting a temperature-independent parameter $\chi_0$ accounting for the core and conduction electrons as well as for a small contribution due to the sample holder, which is sample and direction dependent:
\[ \frac{1}{\chi - \chi_0} = \frac{1}{\chi_{CF}} - \lambda. \]
$\chi_0$ was obtained by adjusting the susceptibility to the calculated free-ion value at $T = 300$~K and takes values in the range between $-2\cdot 10^{-3}$ emu/mol and $-4\cdot 10^{-4}$ emu/mol. A rough estimate of the diamagnetism of the core electrons based on a superposition of the values given by Haberditzl\cite{haberditzl1968} for the single ions results in a lower boundary for the  core contribution  $\chi_{core} \approx 1.25 \cdot 10^{-3}$~emu/mol, rendering this contribution quit sizable at high temperatures. The data shown in Fig.~\ref{fritsch_f6} agree very well for both field directions with the calculations using an isotropic mean-field constant $\lambda = - 8$~mol/emu (in contrast to the direction-dependent parameters differing by a factor of $5$ found in ref.~\onlinecite{joshi2010}). The Curie-Weiss temperature $\theta_{CW}$ is obtained from the data via
\begin{eqnarray*}
\chi - \chi_0 &=  \frac{\chi_{CF}}{1-\lambda \chi_{CF}} = \frac{\chi_{CF}T}{T - \lambda \chi_{CF} T} \equiv \frac{C}{T - \Theta_{CW}}
\end{eqnarray*}
using the Curie constant $C = \chi_{CF}T = 0.807$ emu/mol\,K for Ce$^{3+}$ yielding  $\theta_{CW} = \chi_{CF} T \cdot \lambda \approx -6.5$~K, in line with the overall antiferromagnetic order in this system.


\begin{figure}
\includegraphics[clip,width=0.49\linewidth]{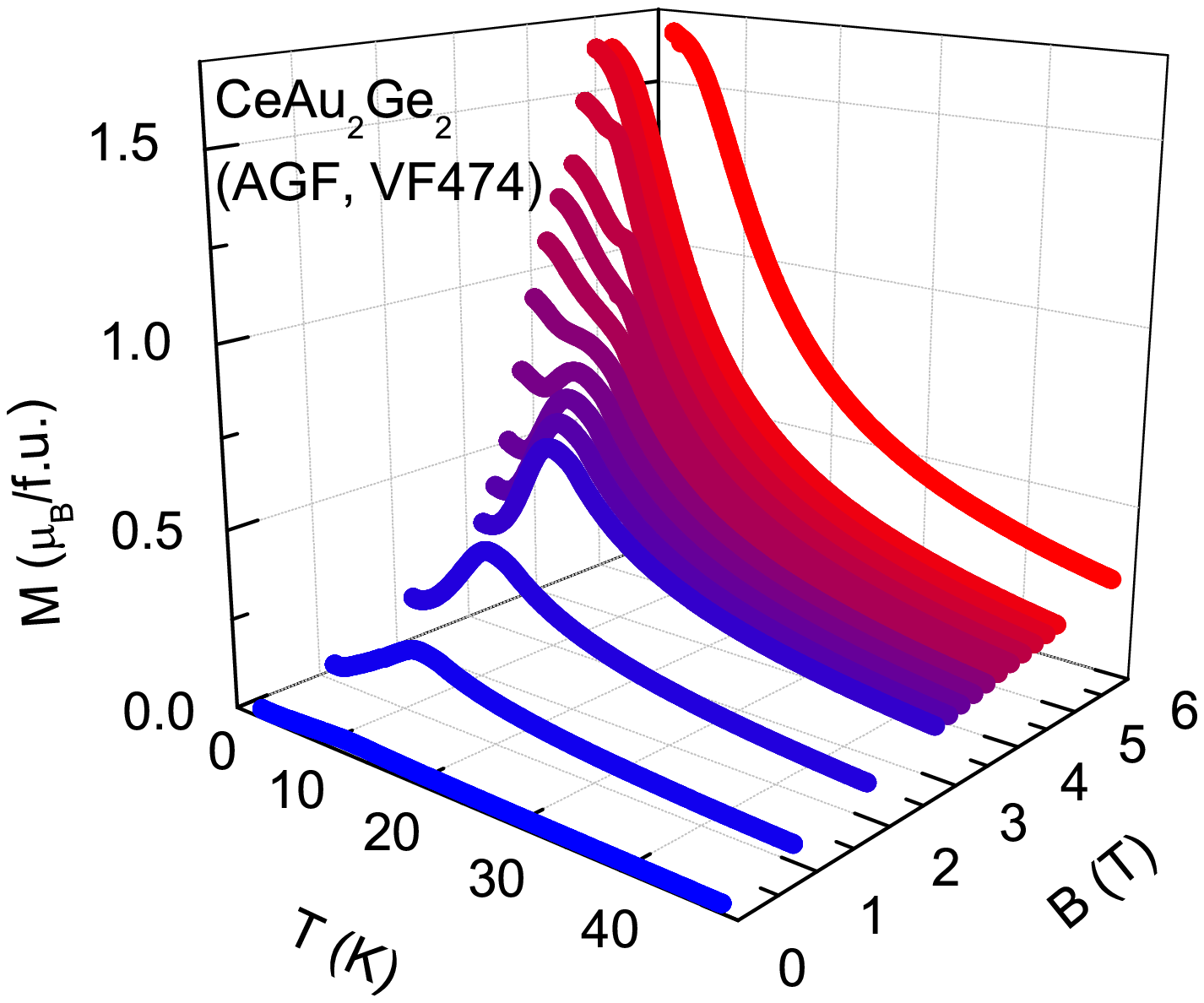}
\includegraphics[clip,width=0.49\linewidth]{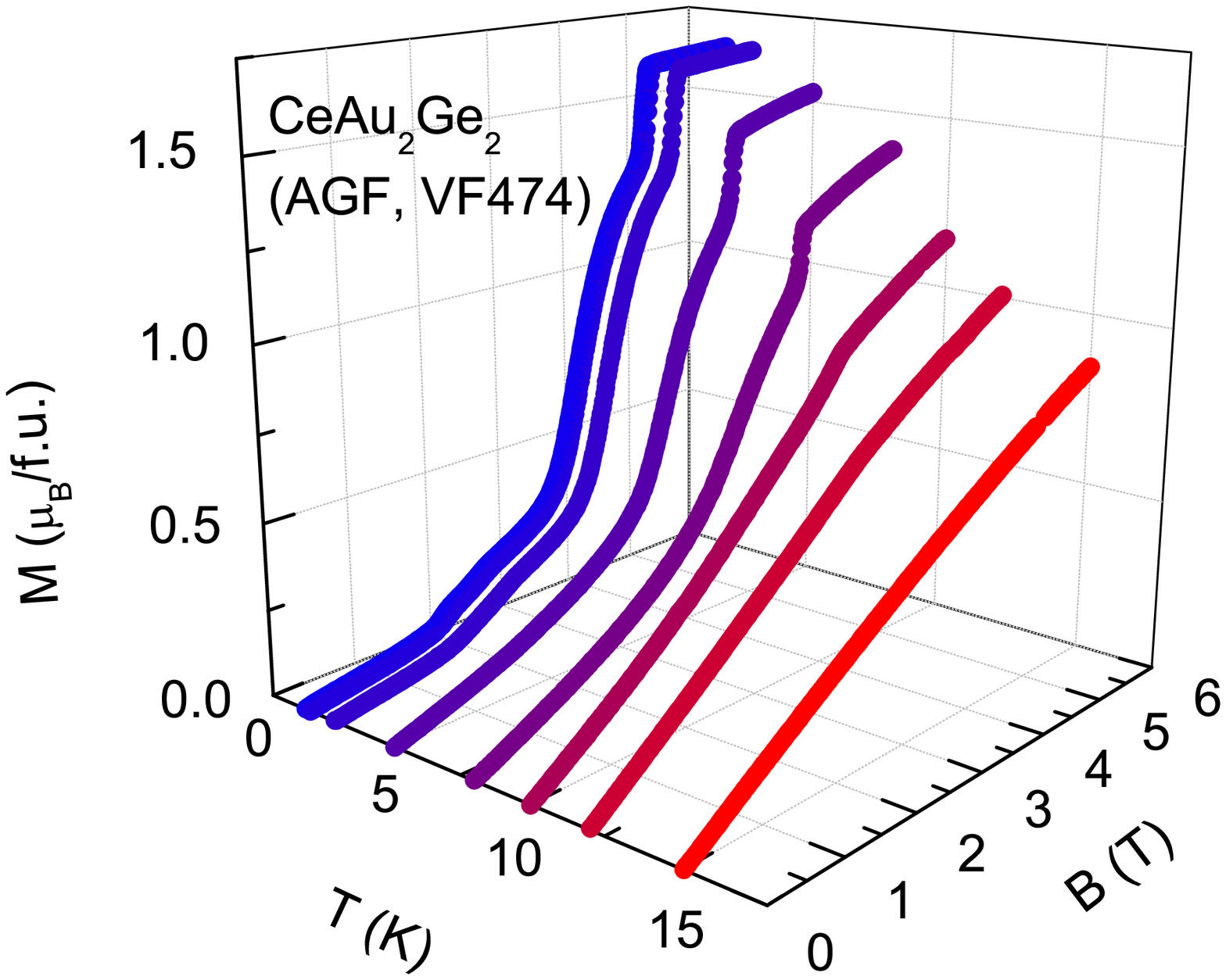}
\caption{(Color online) Left: Magnetization $M$ vs. temperature $T$ in different magnetic fields $\mathbf{B} \| \mathbf{c}$. Right: $M$ vs. external magnetic field $\mathbf{B} \| \mathbf{c}$ at different temperatures $T$.}\label{fritsch_f7}
\end{figure}

\begin{figure}
\includegraphics[clip,width=0.66\linewidth]{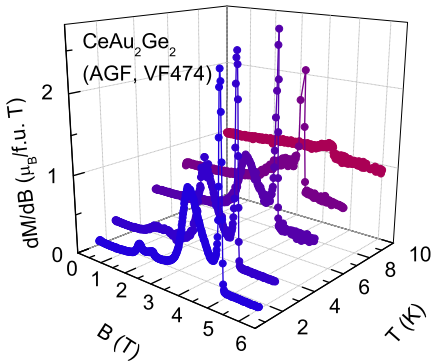}
\caption{(Color online) Magnetic susceptibility $\chi = dM/dB$ of  CeAu$_2$Ge$_2$ versus field $\mathbf{B}\|\mathbf{c}$ at $1.6$, $2.6$, $5$, $7$ and $10$~K.} \label{fritsch_f8}
\end{figure}

In order to characterize the complex magnetic behavior of AGF CeAu$_2$Ge$_2$ in more detail we performed measurements of the magnetization on one sample at various temperatures and fields. In the left-hand panel of Fig.~\ref{fritsch_f7}, temperature sweeps of the magnetization $M(T)$ between $2$ and $50$~K in different magnetic fields are displayed. As expected for antiferromagnetic order, the maximum in $M(T)$ moves to lower temperatures with increasing field. In high fields an additional anomaly develops at higher temperatures. By close inspection of the $M(T)$ data, this anomaly can be traced to low fields, it is even visible in $0.1$~T, as marked by arrows in Fig.~\ref{fritsch_f3}. The right-hand panel of Fig.~\ref{fritsch_f7} shows magnetic field sweeps (with increasing field only) at different temperatures. As already seen in Fig.~\ref{fritsch_f3}, the positions of the metamagnetic transitions in the AGF samples at low temperatures are slightly sample dependent. Furthermore, the values of $M$ at the `plateaus' between the transitions differ strongly for the two samples. For a better characterization of the metamagnetic transitions the magnetic susceptibility $dM/dB$ vs. $B$ is plotted for one sample in Fig.~\ref{fritsch_f8}. With decreasing temperature the metamagnetic transitions, visible as maxima in $dM/dB$ vs $B$,  become sharper. At $T = 1.6$~K four transitions can be clearly identified.

\section{Phase diagram and Discussion}
\begin{figure}
\includegraphics[clip,width=0.48\linewidth]{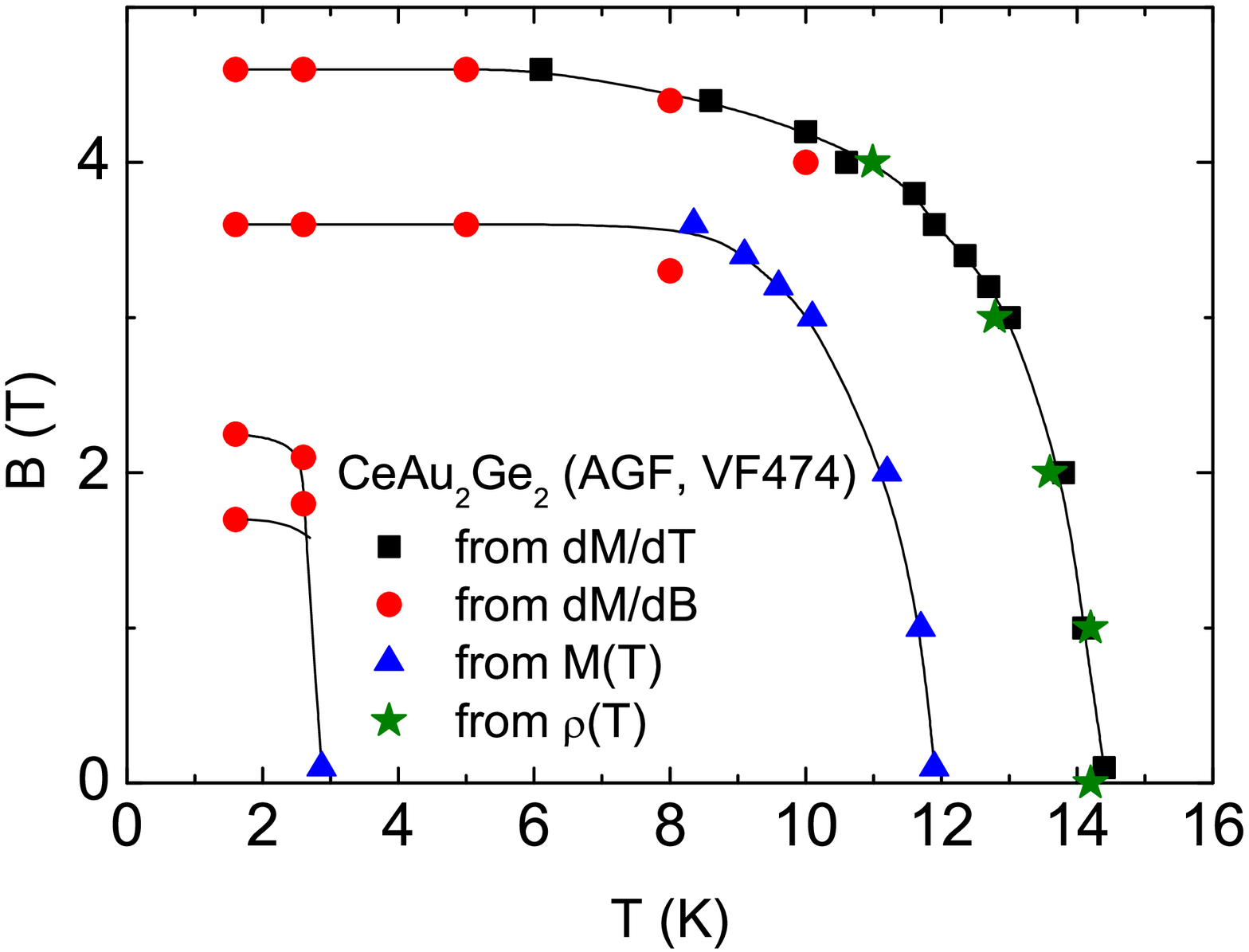}
\includegraphics[clip,width=0.48\linewidth]{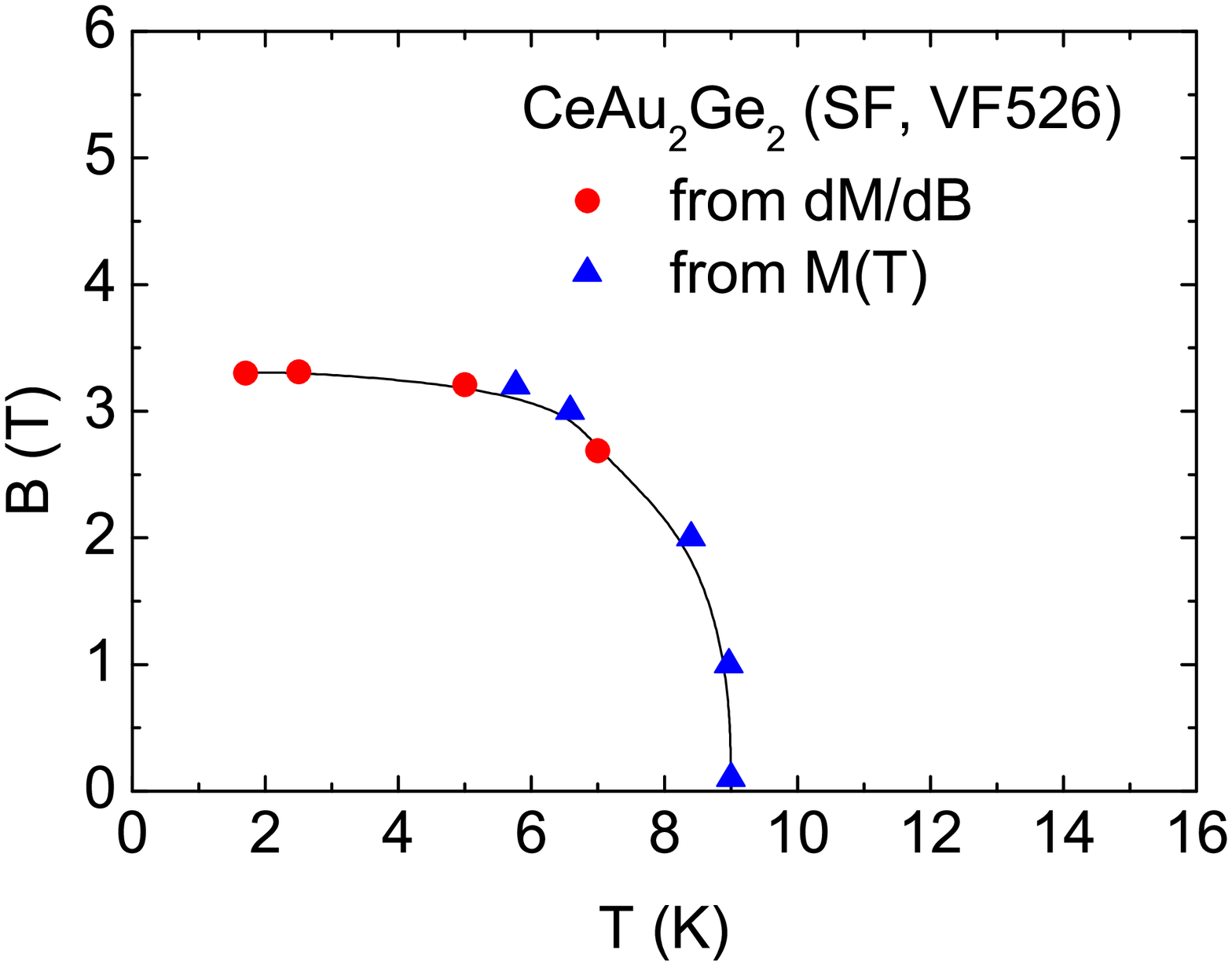}
\caption{(Color online) $B$-$T$ phase diagram of AGF (left) and SF (right) CeAu$_2$Ge$_2$ with the magnetic field aligned parallel to the $c$-axis.} \label{phd}
\end{figure}

From our data we construct $B$-$T$ phase diagrams of AGF and SF CeAu$_2$Ge$_2$ for $\mathbf{B \| c}$. The results are shown in Fig.~\ref{phd}.
For the SF samples (right-hand panel) we have a single transition with anomalous hysteresis behavior separating the antiferromagnetic from the paramagnetic state. For the AGF sample the phase diagram (left-hand panel) reveals the existence of several magnetically ordered phases. The introduction of Sn atoms  through flux apparently prevents the formation of the different phases, indicating a simpler ground state. The introduction of Sn atoms in the SF samples yields a small peak in the $M(T)$ data for $\mathbf{B} \perp \mathbf{c}$ (see Fig.~\ref{fritsch_f3}) and, in addition, a clear signature of a metamagnetic transition at $B^{\ast} = 32$~T for $\mathbf{B} \perp \mathbf{c}$. Both these features for $\mathbf{B} \perp \mathbf{c}$ are absent for the AGF samples. Thus we conclude that CeAu$_2$Ge$_2$ may be viewed as  a system where a comparatively simple order is stabilized by disorder: the introduction of disorder in the form of Sn impurities in the SF samples  yields a simpler magnetic structure, while the AGF samples exhibit a complex magnetic order.

The most frequent reason for impurities leading to a simpler magnetic structure is frustration, which could also account for the magnetization plateaus.\cite{ueda2005}  However, frustration in our system might be excluded by way of the mean-field calculations of the magnetization described above, since the resulting Curie-Weiss $\Theta_{CW}$ temperature is of the same order of magnitude as the magnetic transition temperature $T_N$. Magnetization plateaus have been observed in   isostructural TbNi$_2$Ge$_2$ (Ref. \onlinecite{budko1999}), in some rare-earth monopnictides \cite{rossat1977}, in PrGa$_2$ (Ref.~\onlinecite{ball1993}), and recently in CeCoGe$_3$\cite{thamizhavel2005}. The strong magnetic anisotropy in all these systems confines the magnetic moments to the $(001)$ direction, resulting in Ising-like behavior. In cubic CeSb the magnetization plateaus are assigned to fractions of the expected saturation moment of $2.1~\mu_B$.\cite{rossat1977} The magnetization plateaus and the complex phase diagram are explained by considering long-range interactions resulting in the periodic stacking of alternating ferromagnetic planes, which are successively aligned along the external field with increasing field.\cite{rossat1977,date1988}

A relatively transparent model to arrive at a complex magnetic phase diagram is the anisotropic next-nearest neighbor Ising (ANNNI) model considered theoretically by Bak and von Boehm \cite{boehm1979,bak1980,bak1986}, as well as Fisher and Selke.\cite{fisher1980,selke1988} The ANNNI model can explain the complex phase diagram of CeSb with many incommensurate phases nicknamed `devil's staircase'. We will tentatively discuss our results within this model.

The ANNNI model assumes ferromagnetic nearest-neighbor in-plane interactions and antiferromagnetic next-nearest-neighbor interactions in the perpendicular direction, which would be in agreement with the magnetic order in CeAu$_2$Ge$_2$ proposed by Loidl {\it et al.}\cite{loidl1992}. The  competing interactions then yield different long-range periodic magnetic structures in the $B$-$T$ phase space. In CeAu$_2$Ge$_2$ the magnetization plateaus cannot be mapped unambiguously to fractions of the saturation moment, as clearly seen from their sample dependence. Impurities in the sample are of course able to weaken the long-range interactions, reducing the steps in the magnetization to a small number, as observed in the AGF samples. In the SF samples the amount of impurities is significantly higher, destroying the long-range interactions almost completely and thus yielding a simple antiferromagnetic phase diagram.

For a more quantitative comparison of the ANNNI model with the measured data a mean-field calculation with the two interaction constants, corresponding to the ferromagnetic and the antiferromagnetic interaction of the model, would be necessary. However, presently no experimental data providing values for these interaction constants are available. Furthermore, since the magnetization plateaus are sample dependent, we assume that the interaction constants also are highly sensitive to the sample composition and the local microscopic environment, making a more quantitative analysis difficult.

It should be pointed out that the ANNNI model treats only spin degrees  of freedom and is thus most suitable for insulating  or semiconducting systems. In our case of metallic CeAu$_2$Ge$_2$ no features of a Kondo effect are seen in the magnetic properties. Hence, the conduction electrons do not seem to play a dominant role, so that the features observed here may be still discussed in terms of the ANNNI model. Of course the conduction electrons mediate the magnetic RKKY interactions. We note that weak features observed \cite{sereni2010} in $dM/dB$ in Ce$_2$Pd$_2$Sn, clearly visible in $dM^2/dB^2$, were interpreted  in terms of the Shastry-Sutherland model \cite{shastry1981} that supports decoupled triplet excitations in insulating magnets.

\section{Conclusion}
In conclusion, our experiments on single-crystalline CeAu$_2$Ge$_2$ grown from Au-Ge flux (AGF) or Sn flux (SF) show that the magnetic properties of this system depend strongly on the flux employed. The unusually high magnetic anisotropy for Ce compounds observed in this system, which is preserved at high temperatures and in high fields, is shown to be caused by the crystal electric fields. While AGF samples show a complex magnetic order with several phases for the magnetic field along the easy direction, the SF samples undergo only a single (hysteretic) spin-flop transition to the paramagnetic state. The phase diagram constructed from magnetization data for the AGF samples can be explained qualitatively within the ANNNI model. The SF samples, containing a considerable amount of Sn impurities, yield a simpler magnetic order. Thus CeAu$_2$Ge$_2$ can be viewed as one of the rare systems where the introduction of disorder by impurities leads to a "less disordered" ground state.

\begin{acknowledgments}
We thank J. Sereni, M. B. Maple, P. Coleman and C.-L. Huang for helpful discussions. This work was supported by the Deutsche Forschungsgemeinschaft through FOR 960 and partly by EuroMagNET II under EU contract number 228043.
\end{acknowledgments}

\end{document}